\documentclass[sigconf]{acmart}
\AtBeginDocument{%
  }

\setcopyright{acmlicensed}
\copyrightyear{2026}
\acmYear{2026}
\acmDOI{XXXXXXX.XXXXXXX}

\acmConference[CCS '26]{33rd ACM Conference on Computer and Communications Security}{November 15--19, 2026}{The Hague, The Netherlands}

\acmISBN{978-1-4503-XXXX-X/2026/11}

\usepackage{booktabs}
\usepackage{tikz}
\usepackage{balance}
\usetikzlibrary{arrows.meta,positioning,calc,fit}

\copyrightyear{2026}
\acmYear{2026}
\setcopyright{cc}
\setcctype{by}
\acmConference[CCS '26]{Proceedings of the 2026 ACM SIGSAC Conference on Computer and Communications Security}{November 15--19, 2026}{The Hague, Netherlands}
\acmBooktitle{Proceedings of the 2026 ACM SIGSAC Conference on Computer and Communications Security (CCS '26), November 15--19, 2026, The Hague, Netherlands}
\acmDOI{10.1145/3830454.3846449}
\acmISBN{979-8-4007-2871-6/2026/11}

\begin{document}

\title{Poster: FedWM-Guard: Thwarting Imagination Poisoning in Federated World Model-based Autonomous Driving}
\renewcommand{\shorttitle}{Poster: FedWM-Guard: Thwarting Imagination Poisoning in Federated WM-AD}

\author{Sheng Liu}

\affiliation{%
  \institution{Networked Systems Security Group (NSS)}
  \institution{KTH Royal Institute of Technology}
  \city{Stockholm}
  \country{Sweden}
}
\email{shengliu@kth.se}

\author{Panos Papadimitratos}
\affiliation{%
  \institution{Networked Systems Security Group (NSS)}
  \institution{KTH Royal Institute of Technology}
  \city{Stockholm}
  \country{Sweden}
}
\email{papadim@kth.se}

\renewcommand{\shortauthors}{Sheng Liu and Panos Papadimitratos}

\begin{abstract}
Federated learning (FL) can improve world model (WM)-based autonomous driving (AD) without centralizing raw private vehicle data, but it also turns model aggregation into a safety-critical integrity boundary. We introduce a new threat in federated WM-AD, namely \emph{imagination poisoning}: compromised vehicles submit bounded WM updates that preserve benign short-horizon predictions yet corrupt long-horizon rollouts (e.g., trigger-conditioned) during training, thereby misleading a downstream planner. We present \emph{FedWM-Guard}, to the best of our knowledge, the first defense to characterize planner-facing rollouts in federated WM-AD, screen authenticated updates in hidden-canary scenarios, audit predicted futures against later observations, and invoke a WM-independent safety shield when persistent inconsistency is detected. Unlike parameter-space defenses, it scores what an update makes the model \emph{imagine}, not only how the update looks. We also outline how we plan to evaluate it under non-IID (not independent and identically distributed) data, adaptive attacks, and benign distribution shift. This work highlights an unexplored domain, federated WM-AD, and its threat surface and potential countermeasures.
\end{abstract}


\maketitle

\section{Introduction}

Advanced autonomous driving (AD) systems should not only recognize or understand the current scene, but also anticipate how it may evolve under different actions. World models (WMs) can enrich AD solutions with this critical capability by rolling the scene forward and exposing predicted futures to the planner~\cite{hafner2025dreamer,hu2023gaia}. Once these predictions enter the planning loop, however, their integrity becomes safety-critical. A small error in the current latent state may seem harmless, yet it can grow over several imagined steps and ultimately change the trajectory selected by the vehicle.

Maintaining an accurate WM requires continual learning from massive amounts of vehicle data, as no single vehicle observes more than a narrow slice of the driving distribution. Moreover, many of the events that matter most, e.g., unusual construction layouts, atypical road-user behavior, and near-misses, are rare and distributed across all vehicles. Centralizing the corresponding camera, LiDAR, and location logs would improve WM performance, but it is often impractical because of privacy, legal, and operational constraints. Federated learning (FL) offers an attractive alternative: vehicles train WMs locally and contribute model updates without uploading their raw driving data~\cite{mcmahan2017fedavg}.

This convenience comes with a new security problem: aggregation combines updates from vehicles that may be authenticated, yet remain physically exposed, compromised, or simply highly heterogeneous. An attacker does not need to make the global model fail in an obvious way: a poisoned update may leave normal short-horizon predictions largely unchanged while altering only the long-horizon rollout associated with a particular route, motion pattern, or trigger. Since the planner acts on those imagined futures, a subtle semantic error can become a collision risk even when validation loss and parameter-space anomaly checks appear normal. We refer to this threat as \emph{imagination poisoning}.

Existing research does not fully address this challenge in the intersection of FL, WMs, and AD. Few studies investigate attacks on WMs, and they focus mainly on inference-time inputs or internal dynamics~\cite{guo2026dreamwrong,zhang2026arb4wm,chen2026trusted}. Robust FL defenses typically examine update magnitude, direction, coordinate statistics, or similarity to a trusted reference~\cite{cao2021fltrust,nguyen2022flame,liu2025safeguarding,liu2026defend}. These defenses ask whether an update looks suspicious in parameter space, but not whether it changes the safety-relevant futures generated by the model. This gap leads to our central questions: \emph{can compromised vehicles poison federated WM-AD through bounded, individually plausible updates? Can a countermeasure detect such influence by examining the futures those updates induce?}

\noindent\textbf{Contributions.}
To the best of our knowledge, we are the first to investigate federated WM-based AD and formalize a threat model in which compromised clients poison planner-facing future rollouts while preserving benign short-horizon behavior. Moreover, we introduce FedWM-Guard, a defense architecture that screens client updates according to the \emph{futures they induce}, audits deployed rollouts against independently observed evidence, and feeds verified audit outcomes into subsequent screening rounds, thereby closing the federated lifecycle. We also outline an evaluation plan that specifies
the evidence required before making empirical claims about the defense.

\section{System Model}
\noindent\textbf{WM.}
We consider a latent dynamics model for closed-loop planning. Historical observations, $o_{t-K+1:t}$, e.g., camera/LiDAR, ego state, and map context, together with optional conditioning, $c_t$, are encoded into a latent state $z_t$, which is rolled forward under a candidate action sequence $u_{t:t+H-1}$. Prediction heads decode the imagined rollout into planner-facing future representations,
\begin{equation}
\begin{aligned}
z_t &= E_{\theta}(o_{t-K+1:t}, c_t), &
\hat z^{u}_{t+h} &= F_{\theta}(\hat z^{u}_{t+h-1}, u_{t+h-1}), \\
\hat Y^{u}_{t+h} &= G_{\theta}(\hat z^{u}_{t+h}), &
a_t &= \pi\big(\{\hat Y^{u}_{t+1:t+H}\}_{u\in\mathcal{U}}\big)
\end{aligned}
\label{eq:wm}
\end{equation}
with $\hat z^{u}_t=z_t$, $h=1,\dots,H$, and $u=u_{t:t+H-1}\in\mathcal{U}$ a candidate action sequence: the planner scores $\mathcal{U}$ and executes only the first action $a_t$ of the selected one, whose superscript we omit. Here $\hat Y$ denotes the planner-facing representation: future occupancy, object states, BEV semantics, or trajectories.

\noindent\textbf{Federated training.}
Each participating vehicle updates the trainable WM dynamics based on local data and submits an update $\Delta_i^r$ in round $r$. The server aggregates local updates,
\begin{equation}
\theta^{r+1}=
\operatorname{Agg}
(
\theta^r,\Delta^r_1,\ldots,\Delta^r_n
)
\label{eq:aggregation}
\end{equation}
where $n$ is the number of participants. 

\begin{figure}[t]
\centering
\begin{tikzpicture}[
  font=\tiny,
  box/.style={draw,rounded corners=1pt,align=center,inner sep=1.8pt,fill=white,
              minimum height=5.2mm,text width=14mm},
  guard/.style={box,very thick,fill=gray!12},
  ev/.style={box,fill=gray!4},
  atk/.style={draw,dashed,rounded corners=1pt,fill=gray!6,align=center,
              inner sep=1.8pt,text width=11mm},
  arr/.style={-{Stealth[length=1.3mm]},semithick},
  dal/.style={-{Stealth[length=1.3mm]},semithick,dashed},
  lbl/.style={font=\tiny,inner sep=0.8pt},
  node distance=3.5mm and 8mm]
\node[box] (logs) {Vehicle logs\\$o,a,c$};
\node[box, below=of logs, xshift=1.3mm,  yshift=1.3mm,  draw=black!40] (s2) {\vphantom{Lg}\\\vphantom{Lg}};
\node[box, below=of logs, xshift=0.65mm, yshift=0.65mm, draw=black!40] (s1) {\vphantom{Lg}\\\vphantom{Lg}};
\node[box, below=of logs]    (local) {Local dynamics\\training};
\node[guard, below=of local] (d1)    {D1: identity \&\\influence bounds};
\node[guard, below=of d1]    (d2)    {D2: candidate\\canary rollouts};
\node[box,  below=of d2]     (agg)   {Weighted\\aggregation};
\node[atk,  left=3mm of local] (poison) {O1/O2 poisoning\\(via C1/C2)};
\node[lbl,  above=0.6mm of logs] {\textbf{Training}};
\node[guard, right=of d2] (d3)     {D3: imagination--\\reality audit};
\node[ev,    below=of d3] (sensor) {Onboard sensors\\(independent)};
\node[guard, right=of d3] (d4) {D4: independent\\shield};
\node[box] (wm)      at (d4|-logs)  {Global world\\model $\theta$};
\node[box] (planner) at (d4|-local) {Fixed planner};
\node[box] (control) at (d4|-agg)   {Control};
\node[lbl, above=0.6mm of wm] {\textbf{Deployment}};
\draw[arr] (logs)--(local);
\draw[arr] (local)--node[lbl,right=0.3mm]{$\Delta_i^r$}(d1);
\draw[arr] (d1)--(d2);
\draw[arr] (d2)--(agg);
\draw[dal] (poison.east)--(local.west);
\draw[dotted,semithick,black!55]
   ([xshift=-2mm,yshift=1.75mm]d1.north west)--([xshift=2mm,yshift=1.75mm]d1.north east);
\node[lbl,black!55,anchor=west,font=\tiny\itshape]
   at ([xshift=3mm,yshift=1.75mm]d1.north east) {client $\vert$ server};
\draw[arr] (wm)--(planner);
\draw[arr] (planner)--(d4);
\draw[arr] (d4)--(control);
\draw[arr] (wm.west) -| node[lbl,pos=0.62,left=0.3mm]{$\hat Y$} (d3.north);
\draw[arr] (sensor)--node[lbl,right=0.3mm]{$\tilde Y$}(d3);
\draw[dal] ([yshift=1.1mm]d3.east)--node[lbl,above=0.1mm]{alarm}([yshift=1.1mm]d4.west);
\draw[arr] (sensor.east)--++(2.2mm,0)|-([yshift=-1.4mm]d4.west);
\draw[dal] (d3.west)--(d2.east);
\node[lbl,align=center,fill=white,inner sep=0.5pt]
   at ($(d2.east)!0.5!(d3.west)$) {audit\\feedback};
\coordinate (lane) at ([xshift=2.8mm]wm.east);
\draw[arr] (agg.south)--++(0,-3.5mm)-|(lane)--(wm.east);
\node[lbl,fill=white,inner sep=0.5pt] at ([yshift=-3.5mm]control.south) {broadcast};
\end{tikzpicture}
\caption{FedWM-Guard separates training-time semantic screening (left) from deployment-time audit and containment (right).}
\Description{Three-column block diagram. Left column (federated training): vehicle logs feed local dynamics training on many clients, shown as a stack of boxes, with an O1/O2 poisoning node attached via C1/C2 capabilities; client updates cross a dotted client-server trust boundary as delta, then pass D1 identity and influence checks and D2 candidate-model canary rollout screening before weighted aggregation, whose result is broadcast back to the deployed model. Right column (on-vehicle deployment): the global WM feeds a fixed planner, then the D4 independent shield, then control. Middle column: a D3 imagination-reality audit compares the WM's rollout Y-hat with an independent onboard-sensor representation Y-tilde; it sits off the control path and only raises an alarm to D4, and it feeds audit outcomes back to D2 screening. The onboard sensors also supply D4 with independent obstacle evidence.}
\label{fig:architecture}
\end{figure}
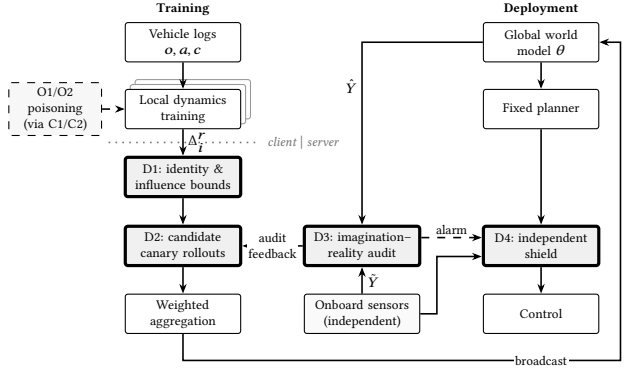

\section{Threat Model}
We assume an authenticated but behaviorally untrusted subset of vehicles. Compromised clients may poison local data or directly optimize updates, coordinate across rounds, and know the aggregation rule and semantic score. Direct actuator takeover and physical sensor destruction are out of scope. 

\noindent\textbf{Adversary goal.}
Writing $\theta'$ for the aggregate~\eqref{eq:aggregation} with the compromised updates injected, and $\hat Y_{t+1:t+H}(\theta')$ for the rollout it induces, the adversary seeks updates that maximize the deviation of $\hat Y_{t+1:t+H}(\theta')$ from the safe reference future (and the resulting planner-level risk) while keeping the benign short-horizon loss within $\epsilon$ of nominal and each update inside the aggregation/screening feasible set. The attack thus preserves normal short-horizon prediction yet steers the long-horizon futures the planner consumes.

\noindent\textbf{Attack objectives.}
\textit{O1: Untargeted rollout degradation.} An input-agnostic update that preserves one-step validation loss but amplifies multi-step compounding error on safety-critical scenes (semantic degradation without any trigger).
\textit{O2: Trigger-conditioned rollout backdoor.} Clean-input behavior is unchanged, but a route, map, motion, V2X, or physical trigger makes an obstacle disappear, shift, or slow incorrectly after several imagined steps.

\noindent\textbf{Evasion capabilities.}
\textit{C1: Collusion.} Several clients split one malicious direction into individually benign-looking updates that pass FLTrust or other robust aggregation-style filters.
\textit{C2: Staleness.} Under semi-asynchronous rounds, stale benign updates legitimately drift further from the global model, widening the tolerance envelope that norm and similarity filters must accept; the attacker times and shapes submissions to hide inside it. In both cases, attackers keep the current scene plausible and act only through the imagined future, exactly what parameter-space FL defenses do not inspect.

\section{FedWM-Guard}
The designed FedWM-Guard includes four checks with different trust assumptions (Figure~\ref{fig:architecture}).

\noindent\textbf{D1: Authenticated, bounded participation.}
Signatures, freshness checks, per-layer clipping, and per-client weight caps prevent spoofing, replay, and single-client domination. D1 establishes provenance and influence bounds, not benign intent.

\noindent\textbf{D2: Rollout-aware aggregation.}
For each surviving update, the server instantiates the candidate model $\theta_i^{r+1}=\theta^r+\eta\Delta_i^r$, where $\eta$ is the server-side step size; and evaluates it on a set of hidden canary scenarios $\mathcal{C}$, extracting four normalized features: short-horizon error $L_{\mathrm{short}}$, multi-step rollout deviation $D_H$ from a trusted (or ensemble) reference, RSS-inspired safety-margin violation $S_{\mathrm{safe}}$~\cite{shalev2017rss}, and ensemble disagreement $U_H$. Each canary fixes the observation context \emph{and} an action sequence, so all candidates roll out under identical conditioning and $D_H$, $U_H$ measure divergence in the learned dynamics. We formulate update screening as benign-versus-poisoned classification via the calibrated statistic,
\begin{equation}
V_i=\mathbf{w}^{\top}\phi_i,\qquad
\phi_i=\big(L_{\mathrm{short}},D_H,S_{\mathrm{safe}},U_H\big)
\label{eq:score}
\end{equation}
whose weights $\mathbf{w}$ can be learned by logistic regression on held-out benign and synthetically poisoned updates, with threshold $\tau$ set to a target false-positive budget. Survivors are aggregated by a soft trust-weighted (Gibbs) rule,
\begin{equation}
q_i \propto
\max\{0,\cos(\Delta_i,\Delta_{\mathrm{root}})\}\,
\exp(-V_i/T)
\label{eq:weight}
\end{equation}
combining directional trust~\cite{cao2021fltrust} with \emph{semantic} trust from rollout screening; $T$ sets the sharpness of the down-weighting, while hard rejection is governed separately by $\tau$. Thus $V_i$ is a \emph{calibrated screening statistic}, not a certified robust-aggregation rule, and the linear detector can be replaced by nonlinear models over $\phi_i$. The server then evaluates the tentative global model on the same canary suite before committing the round, catching coalitions (C1) whose updates are individually benign.

\noindent\textbf{D3: Imagination-reality audit.}
At deployment, the stored latent state $z_t$ is replayed under the logged realized action sequence $a_{t:t+H-1}$, using the model version that deployed at time $t$. The resulting compact planner-facing projection $\Pi(\hat Y^{\mathrm{rep}}_{t+h})$ is
compared with $\Pi(\tilde Y_{t+h})$, where $\tilde Y_{t+h}$ derived directly from onboard sensor observations, independently of the audited WM. A persistent prediction-reality inconsistency raises a risk alarm, 
\begin{equation}\label{eq:audit}
e_t=\sum\nolimits_{h=1}^{H}\omega_h\,
d\!\left(\Pi(\hat Y^{\mathrm{rep}}_{t+h}), \Pi(\tilde Y_{t+h})\right)
\end{equation}
where $d(\cdot,\cdot)$ is a task-specific rollout discrepancy. Persistence and uncertainty gating suppress transient noise, while a sustained residual is treated as a risk signal.

\noindent\textbf{D4: Independent containment.}
When risk remains high, the planner's output is restricted by a fallback controller that does not share the attacked latent dynamics. It uses ego kinematics, conservative obstacle envelopes, and RSS-inspired constraints~\cite{shalev2017rss}. The fallback controller derives these obstacle envelopes from onboard sensors through an independent perception pipeline (the same evidence source used by D3), not from the audited WM. D4 limits impact even when poisoning evades training-time checks.

\smallskip\noindent\textbf{Closing the loop.}
Deployment evidence is not discarded locally: persistent alarms corroborated across vehicles in a region are reported to the server as a round- and region-level integrity signal. After server-side verification, a descriptor of the offending scenario is used to add an equivalent scenario (e.g., reconstructed in simulation or retrieved from the server's root set) to the canary set $\mathcal{C}$, and updates from the implicated rounds are flagged for re-screening. Runtime audits thus tighten next-round screening rather than only degrading one vehicle, making the defense a federated \emph{lifecycle}.

\section{Evaluation Plan}

We plan to evaluate a compact DreamerV3-style dynamics module~\cite{hafner2025dreamer} federated across heterogeneous CARLA clients~\cite{dosovitskiy2017carla} (distinct maps, weather, and traffic as non-IID sources), with encoder, planner, and shield fixed. Here $\hat Y$ is $H$-step BEV occupancy and $\pi$ scores trajectories against the \emph{predicted} occupancy, while D3 compares it against \emph{independent} occupancy from CARLA ground truth perturbed by a perception noise model; this makes the D2 score and D3 residual well-defined. On-vehicle feasibility assumes idle-time local fine-tuning, not in-drive or from-scratch training. Table~\ref{tab:evaluation} defines the evidence required before we make empirical claims.

\begin{table}[t]
\caption{Plan for evaluation.}
\label{tab:evaluation}
\scriptsize
\renewcommand{\arraystretch}{1.25}
\begin{tabular}{@{}p{0.22\columnwidth}p{0.72\columnwidth}@{}}
\toprule
Question & Factors and evidence \\
\midrule
Attack effectiveness &
Vary malicious ratios, triggers, budgets, non-IID splits, canary size/coverage, collusion, and staleness. Report O1/O2 rollout error, ASR, short-horizon utility, and closed-loop risk. \\

Detection &
Vary weather, traffic, routes, sensor noise, and unseen domains. Report AUROC/AUPRC, TPR at 1\%/5\% FPR, delay, and benign rejection. \\

Safety benefit &
Ablate D1--D4 over fixed routes and $\geq 3$ seeds. Report route completion, collisions/infractions per km, min TTC, and confidence intervals. \\

Cost &
Vary client count, participation, and horizon.
Report aggregation latency, canary cost, communication, and runtime overhead. \\
\bottomrule
\end{tabular}

\vspace{1.0pt}\par
{\scriptsize\itshape Comparators: FLTrust, FLAME, parameter-only
screening, current-frame checking, and ablations.\par}
\end{table}

\noindent\textbf{Adaptive attack protocol.}
The attacker jointly minimizes clean degradation and defense score while maximizing delayed rollout error and planner risk (full knowledge of Eqs.~(\ref{eq:score})-(\ref{eq:weight})), optionally layering C1/C2. Budgets, participation, routes, thresholds, and seeds are fixed before evaluation.

\section{Discussion}

There are several open questions for federated WM-AD security. Which planner-facing representation is interpretable yet independent enough for semantic canaries? How to calibrate audits across heterogeneous benign vehicles without rejecting rare safety-critical clients? Can D2's per-candidate scoring be reconciled with secure aggregation (inspecting gradients), while recovering gradient confidentiality at acceptable cost? How can the counterfactual blind spot in D3 be mitigated?

\section{Conclusion}

Federated WM-AD introduces a new security risk: at model aggregation, a few compromised clients can poison the futures used for planning. FedWM-Guard defends that point by screening updates based on the futures they induce: the question is not whether an update is statistically unusual, but whether it leads the vehicle to imagine an unsafe world.

\bibliographystyle{ACM-Reference-Format}
\bibliography{poster-wm}

\end{document}